\documentclass[twocolumn,english,showpacs]{revtex4}
\usepackage{amsmath,amssymb,dcolumn}
\usepackage{epsfig}
\def\imo{i}
\begin{document}
\title{Looking at the Gregory-Laflamme instability through quasi-normal modes}
\author{R. A. Konoplya}\email{konoplya_roma@yahoo.com} \author{Keiju Murata}\email{murata@tap.scphys.kyoto-u.ac.jp} \author{Jiro Soda} \email{jiro@tap.scphys.kyoto-u.ac.jp}
\affiliation{Department of Physics, Kyoto University, Kyoto 606-8501, Japan}
\author{A. Zhidenko}\email{zhidenko@fma.if.usp.br}
\affiliation{Instituto de F\'{\i}sica, Universidade de S\~{a}o Paulo \\
C.P. 66318, 05315-970, S\~{a}o Paulo-SP, Brazil}
\begin{abstract}
We study evolution of gravitational perturbations of black strings.
It is well known that for all wavenumber less than some threshold value,
the black string is unstable against scalar type of gravitational perturbations, which is named
the Gregory-Laflamme instability. Using numerical methods,
 we find the quasinormal modes and time-domain profiles of the black string perturbations in the stable sector and also show the appearance of the Gregory-Laflamme instability in the time domain. The dependence of the black string quasinormal spectrum and late time tails on such parameters as  the wave vector and the number of extra dimensions is discussed. There is a numerical evidence that in the threshold point of instability the static solution of the wave equation is dominant. For wavenumbers slightly larger than the threshold value, in the region of stability, we see tiny oscillations with very small damping rate. While, for wavenumbers slightly smaller than the threshold value, in the region of the Gregory-Laflamme instability,
 we observe tiny oscillations with very small growth rate.
 We also find the level crossing of imaginary part of quasinormal modes
 between the fundamental mode and the first overtone mode, which accounts for
 the peculiar time domain profiles.
\end{abstract}
\maketitle

\section{Introduction}

Unlike four dimensional Einstein gravity, which allows existence of black holes, higher dimensional theories, such as the brane-world scenarios and string theory, allow existence of a number of "black" objects: higher dimensional black holes, black strings and branes, black rings and saturns and others. In higher than four dimensions we lack the uniqueness theorem, so that stability may be the criteria which will select physical solutions among this variety of solutions. Up to now, we know that higher dimensional Reissner-Nordstr\"om-de Sitter black holes are stable \cite{ZhidenkoNPB} in the Einstein gravity. On the contrary, black holes in Gauss-Bonnet (GB) gravity are unstable for large GB coupling
for $D =5, 6$ \cite{ZhidenkoGB}, where $D$ is the total number of space-time dimensions. Kaluza-Klein black holes with squashed horizon
are stable against lowest zero mode perturbations \cite{Ishihara}. Unlike, Kaluza-Klein black holes, the black
string metric is a solution of the Einstein equations in five or higher dimensional gravity that has a
factorized form consisting of the Tangherlini black hole and an extra flat dimension \cite{Harmark:2007md}. According to the brane-world scenarios, if the matter localised on the brane undergoes gravitational collapse, a black hole with the horizon extended to the transverse extra direction will form. This object looks like a black hole on the brane, but is, in fact, a black string in the full D-dimensional theory.

It is well known that such black strings suffer from the so-called Gregory-Laflamme instability, which is the long wavelength gravitational instability of the scalar type of the metric perturbations \cite{Gregory:1993vy}, \cite{Gregory:1994bj}. The Gregory-Laflamme instability has been intensively studied during the recent decade \cite{Harmark:2007md} and the threshold values of the wave vector $k$ at which the instability appears are known \cite{Hovdebo:2006jy}. In the present paper we are aimed at studying the evolution of linear perturbations of D-dimensional black strings in time and frequency domains. This task is motivated mainly by the two reasons: first to realize what happens on the edge of instability of black strings and how the perturbations will develop in time. Second, the so-called quasinormal modes of a stable black string might be an observational characteristic for the future Large Hadron Collider Experiments, if such objects as black strings exist.

The latter needs some more explanation. In this research, we shall show that,
 if a stable D-dimensional black string is gravitationally perturbed,
it will undergo damped oscillations, called quasinormal ringing, similar to that of a black hole \cite{Kokkotas}. At asymptotically late time, this quasinormal
ringing goes over into the power-law tails. The quasinormal modes and asymptotic tails are very well studied for D-dimensional black holes
\cite{Konoplya:2004xx} and for black holes localised on the brane \cite{Kanti:2005xa}. We
find  here that the quasinormal
ringing of black strings has a number of differences from that of D-dimensional black holes \cite{Konoplya:2004xx}, especially when approaching the edge of instability.
In particular, we find a numerical evidence that in the threshold point of instability, the static solution of the wave equation is dominant.  In the region of the stability, for
$k$ slightly larger than the threshold value,
we see modes with tiny oscillation frequencies and damping rates.
In the region of the   Gregory-Laflamme instability, for $k$
slightly smaller than threshold value,  we observe tiny oscillations and instability growth.
We also find the level crossing of imaginary part of quasinormal modes
 between the fundamental mode and the first overtone mode.

The paper is organized as follows. Sec II gives basic formulas for the black string perturbations and the wave equation for the scalar type
of gravitational perturbations. Sec III analyses quasinormal modes of black strings in frequency domain with the help of the Frobenius method, and in
time domain by the Gundlah-Price-Pullin method.
 We also discuss features of the  Gregory-Laflamme instability
  using the obtained results.
 The final section is devoted to the conclusion.

\section{Basic formulae}

In this section, we shall briefly review the results of the paper \cite{Hovdebo:2006jy}, where the wave equation for the scalar type of gravitational perturbations were obtained. This wave equation will be our starting point for numerical investigation.

For the static string in $D=n+4$ space-time dimensions, the background metric can be written as
\begin{equation}
ds^2=g_{\mu\nu}dx^\mu dx^\nu=-f(r)dt^2+\frac{dr^2}{f(r)}+r^2d\Omega^2_{n+1}+dz^2,
\end{equation}
where
$$f(r)=1-\left(\frac{r_+}{r}\right)^n,$$
and $d\Omega^2_{n+1}$ is the metric on a unit $(n+1)$-sphere. Various properties of black string have been extensively studied
in recent years and we refer a reader to the papers \cite{Liu:2008ds}-\cite{Kanno:2003yu} for more detailed information on black strings.

The $z$-direction is periodically identified by the relation $z=z+2\pi R$. We study perturbations of an $(n+1)$-spherically symmetric solution with the Killing vector in $z$-direction. Therefore, we can write perturbations in the form
$$\delta g_{\mu\nu}=e^{\imo kz}a_{\mu\nu}(t,r), \qquad k=\frac{m}{R}, \quad m\in\mathbb{Z}.$$
The perturbed vacuum Einstein equations have the form
\begin{equation}
\delta R_{\mu \nu} = 0
\end{equation}
The perturbations can be reduced to the form, where the only non-vanishing components of $a_{\mu\nu}$ are
$$a_{tt}=h_t, \quad a_{rr}=h_r, \quad a_{zz}=h_z,$$ $$a_{tr}=\dot{h}_v, \quad a_{zr}=-\imo k h_v.$$

The linearized Einstein equations give a set of coupled equations determining the four radial profiles above.
However, we may eliminate $h_v$, $h_r$ and $h_t$ from these equations in order to produce a single second order equation for $h_z$:
\begin{equation}\label{HovdeboMyers}
\ddot{h}_z=f(r)^2h_z''+p(r)h_z'+q(r)h_z,
\end{equation}
where
\begin{eqnarray}\nonumber
p(r)&=&\frac{f(r)^2}{r}\left(1+\frac{n}{f(r)}-\frac{4(2+n)k^2r^2}{2k^2r^2+n(n+1)(r_+/r)^n}\right),\\\nonumber
q(r)&=&-k^2f(r)\frac{2k^2r^2-n(n+3)(r_+/r)^n}{2k^2r^2+n(n+1)(r_+/r)^n}.
\end{eqnarray}

Defining
$$h_z(t,r)=\frac{r^{-(n-1)/2}}{2k^2r^2+n(n+1)(r_+/r)^n}\Psi(t,r),$$
we can reduce the equation (\ref{HovdeboMyers}) to the wave-like equation
\begin{equation}\label{wavelike}
\left(\frac{\partial^2}{\partial t^2}-\frac{\partial^2}{\partial r_\star^2} + V(r)\right)\Psi=0,
\end{equation}
where $dr_\star=\frac{dr}{f(r)}$ is the tortoise coordinate.
Here, the effective potential $V(r)$ is given by
$$V(r)=\frac{f(r)}{4 r^2}\frac{U(r)}{\left(2k^2r^2+n(n+1)(r_+/r)^n\right)^2} \ , $$
where
\begin{eqnarray}\nonumber
&&U(r)=16 k^6 r^6+ 4 k^4 r^4 (n+5) (3f(r)-2 n + 3n f(r))-\\\nonumber&&-4k^2r^2n(n\!+\!1)\left(n(n\!+\!5)+f(r)(2n^2\!+\!7n\!+\!9)\right)\left(\frac{r_+}{r}\right)^n\!\!-
\\\nonumber&&-n^2(n+1)^3\left(f(r)-2n+nf(r)\right)\left(\frac{r_+}{r}\right)^{2n}.
\end{eqnarray}

The above effective potential does not vanish at asymptotic infinity but has an effective "mass" term, containing $k$, at the spatial infinity.

\section{Evolution of perturbations analyzed with the Frobenius method and time domain integration technique}

First of all, let us briefly describe the two methods which we used here: the Frobenius method  (frequency domain) and the Gundlach-Price-Pullin method (time domain).

In time domain, we study the black string ringing using a numerical characteristic integration method \cite{Gundlach:1993tp}, that uses the light-cone variables $u = t - r_\star$ and $v = t + r_\star$. In the characteristic initial value problem, initial data are specified on the two null surfaces $u = u_{0}$ and $v = v_{0}$. The discretization scheme we used, is
\begin{eqnarray}\label{d-uv-eq}
\Psi(N) &=& \Psi(W) + \Psi(E) - \Psi(S) -\\\nonumber&&-\Delta^2\frac{V(W)\Psi(W) + V(E)\Psi(E)}{8} + \mathcal{O}(\Delta^4) \ ,
\end{eqnarray}
where we have used the following definitions for the points: $N =(u + \Delta, v + \Delta)$, $W = (u + \Delta, v)$, $E = (u, v + \Delta)$ and $S = (u,v)$.

In frequency domain we used the well-known Frobenius method \cite{Leaver:1985ax}.
In order to study the QN spectrum in frequency domain, we have separated time and radial coordinates in (\ref{HovdeboMyers})
$$h_z(t,r)=e^{-\imo\omega t}h_\omega(r).$$
Here $h_\omega(r)$ satisfies the quasinormal mode boundary conditions, which are purely ingoing wave at the event horizon and purely outgoing wave at the spatial infinity.
Thus, the appropriate Frobenius series are
\begin{equation}\label{Frobenius}
h_\omega(r)=\left(1 - \frac{r_+}{r}\right)^{-\imo\omega r_+/n} e^{\imo\sqrt{\omega^2-k^2}r}r^{(n+3+\alpha)/2}y(r),
\end{equation}
where $\displaystyle\alpha=\frac{\imo(2\omega^2-k^2)r_+}{\sqrt{\omega^2-k^2}}$ for $n=1$ and $\alpha=0$ for $n>1$.
It is crucial that $y(r)$ must be regular at the event horizon and at the spatial infinity and can be expanded as
$$y(r)=\sum_{i=0}^{\infty}a_i\left(1 - \frac{r_+}{r}\right)^i.$$
Substituting (\ref{Frobenius}) into (\ref{HovdeboMyers}), we have found that the coefficients $a_i$ satisfy a $(3n+5)$-term recurrence relation.
We found the coefficients of the recurrence relation, and then we obtained the equation with the infinite continued fraction, which is algebraic equation with respect to the QN frequency $\omega$. Numerical solutions of this algebraic equation give us the QN spectrum.

\begin{table}
\caption{Fundamental mode ($\omega_0$) found by the time-domain integration, first ($\omega_1$) and second ($\omega_2$) overtones of spherically symmetric black string perturbations ($n=1$) found by the Frobenius method. As $k$ grows, the first overtone decreases its imaginary part, becoming the fundamental mode, reaching purely real frequency (quasi-resonance) at $k\approx0.94$ and then disappearing. The second overtone, as well as the fundamental purely imaginary mode, increases its damping rate with $k$.}\label{n=1.table}
\begin{tabular}{|c|c|c|}
\hline
$k$&$\omega_0$ (t-d)&$\omega_1$ (Frob.)\\
\hline
$0.84$&$+0.011\imo$&$0.802-0.0173\imo$\\
$0.85$&$+0.008\imo$&$0.810-0.0157\imo$\\
$0.86$&$+0.005\imo$&$0.818-0.0140\imo$\\
$0.87$&$+0.002\imo$&$0.827-0.0124\imo$\\
$0.88$&$-0.001\imo$&$0.835-0.0107\imo$\\
$0.89$&$-0.004\imo$&$0.844-0.0090\imo$\\
$0.90$&$-0.007\imo$&$0.852-0.0074\imo$\\
$0.91$&$-0.011\imo$&$0.861-0.0058\imo$\\
$0.92$&$-0.014\imo$&$0.870-0.0042\imo$\\
$0.93$&$-0.017\imo$&$0.879-0.0025\imo$\\
$0.94$&$-0.021\imo$&$0.888-0.0007\imo$\\
\hline
\end{tabular} \qquad \begin{tabular}{|c|c|}
\hline
$k$&$\omega_2$ (Frob.)\\
\hline
$0.8$&$0.361-0.632\imo$\\
$0.9$&$0.373-0.679\imo$\\
$1.0$&$0.378-0.724\imo$\\
$1.1$&$0.374-0.767\imo$\\
$1.2$&$0.364-0.806\imo$\\
$1.3$&$0.347-0.841\imo$\\
$1.4$&$0.324-0.870\imo$\\
$1.5$&$0.294-0.894\imo$\\
\hline
\end{tabular}
\end{table}

Now let us discuss the obtained results for QN modes and time domain profiles.
The Frobenius method for the considered cumbersome potential gives rise to a number of technical difficulties: first, the convergence
of the Frobenius series is rather slow. Second, when searching for the solutions of the algebraic equation in the region close to the threshold of instability,
one needs very good initial guess for $\omega$ to "fall" into the minimum of the continued fraction equation. This can be easily understood.  As we shall
see from the time domain integration, the dominant solution in the threshold point corresponds to some static solution with $\omega =0$, so that nearby
fundamental mode has tiny real and imaginary parts. The Frobenius method naturally requires slow convergence and excellent initial guess for $\omega$ for such small $\omega$.

\begin{figure*}
\includegraphics[width=.5\linewidth]{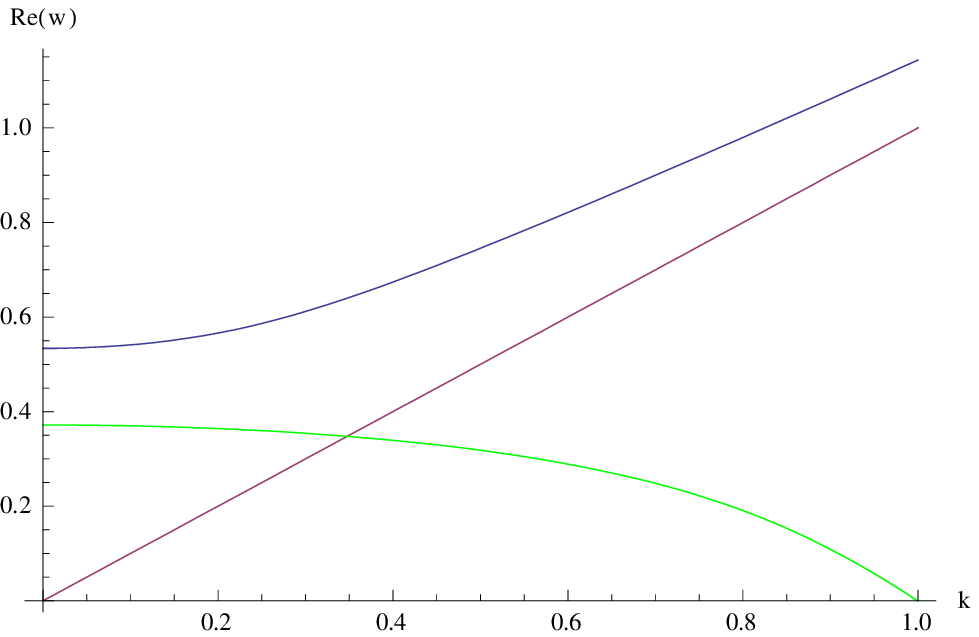}\includegraphics[width=.5\linewidth]{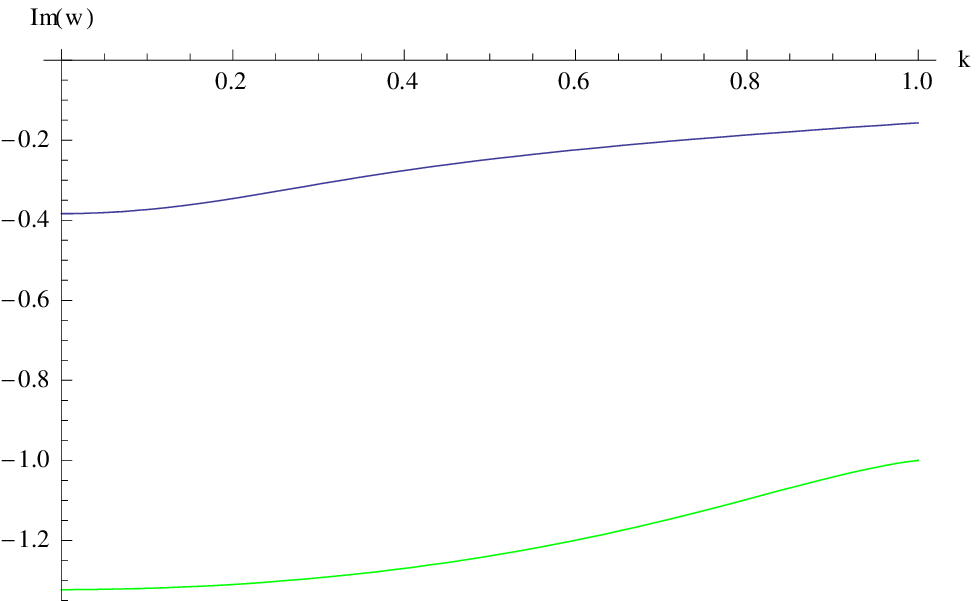}
\caption{Real and imaginary part of first two overtones for $n=2$ black string perturbations as a function of $k$. The first overtone $\omega_1$ (blue) approaches $\omega=k$ (red line). The second overtone $\omega_2$ (green) becomes pure imaginary in stable region ($k\sim1$).}\label{n=2.fig}
\end{figure*}

\begin{table}
\caption{Dominant QNMs of spherically symmetric black string perturbations ($n=2$) found with the
time-domain and the Frobenius method. For $k \geq 1.4$ the $\omega_0$ mode is not dominating anymore and is difficult for detection
by the time domain integration.}
\begin{tabular}{|c|c|c|c|}
\hline
$k$&$\omega_0$(t-d)&$\omega_1$(t-d)&$\omega_1$ (Frobenius)\\
\hline
$1.3$&$-0.012\imo$&$1.418-0.141\imo$&$1.396-0.118\imo$\\
$1.4$&$-0.047\imo$&$1.491-0.099\imo$&$1.483-0.107\imo$\\
$1.5$&$n/a$&$1.584-0.095\imo$&$1.570-0.096\imo$\\
$1.6$&$n/a$&$1.674-0.088\imo$&$1.659-0.086\imo$\\
$1.7$&$n/a$&$1.763-0.081\imo$&$n/a$\\
$1.8$&$n/a$&$1.852-0.077\imo$&$n/a$\\
$1.9$&$n/a$&$1.936-0.054\imo$&$n/a$\\
$2.0$&$n/a$&$2.032-0.046\imo$&$n/a$\\
\hline
\end{tabular}
\end{table}

 In tables I, we have listed
 fundamental mode $\omega_0$, first $\omega_1$ and second $\omega_2$
  overtone of spherically symmetric black string perturbations in the case of $n=1$.
   First, we notice
 the level crossing of imaginary part of quasinormal modes between the fundamental mode
 and the first overtone mode. This level crossing is peculiar to the black strings.
 As $k$ grows, the first overtone $\omega_1$ decreases its imaginary part, becoming the fundamental mode, reaching purely real frequency (quasi-resonance) at $k\approx0.94$ and then disappearing. The second overtone, as well as the fundamental purely imaginary mode, increases its damping rate with $k$.

In table II, we listed dominant QNMs of spherically symmetric black string
perturbations in the case of $n=2$.
 We can see that the fundamental mode near
the threshold of instability has no real part.
Yet, higher overtones have detectable real part.

\begin{table}
\caption{Dominant QNMs of spherically symmetric black string perturbations ($n=3$) found with the time-domain method.
For $k \geq 2$ the $\omega_1$ mode is a dominating one.}
\begin{tabular}{|c|c|c|}
\hline
$k$&$\omega_0$&$\omega_1$\\
\hline
$1.6$&$-0.007\imo$&$1.869-0.214\imo$\\
$1.7$&$-0.043\imo$&$1.930-0.211\imo$\\
$1.8$&$-0.082\imo$&$2.013-0.212\imo$\\
$1.9$&$-0.130\imo$&$2.106-0.185\imo$\\
$2.0$&$-0.250\imo$&$2.223-0.179\imo$\\
$2.1$&$n/a$&$2.309-0.172\imo$ \\
$2.2$&$n/a$&$2.400-0.167\imo$ \\
$2.3$&$n/a$&$2.472-0.161\imo$ \\
\hline
\end{tabular}
\end{table}

In table III, we listed dominant QNMs of spherically symmetric black string
perturbations in the case of $n=3$.
 We can see
 the level crossing of imaginary part of quasinormal modes between the fundamental mode
 and the first overtone mode.

The real part of the second mode ($\omega_1$ in tables I, III) asymptotes to
$k$ at large $k$, while the imaginary part monotonically decreases when $k$ is increasing.
 The third mode does not asymptotes to $k$, but has monotonically decreasing real
and imaginary parts as can be seen from Fig. 1. Because of the risk to "fall" into another overtone, in order to obtain the higher overtones, we had to
start from the D-dimensional black holes with $k=0$, for which the QN frequencies are known \cite{Konoplya:2003dd}, and then to "move" towards
higher $k$ in the Frobenius method with a very small step (see Fig. 1).

\begin{figure}
\includegraphics[width=\linewidth]{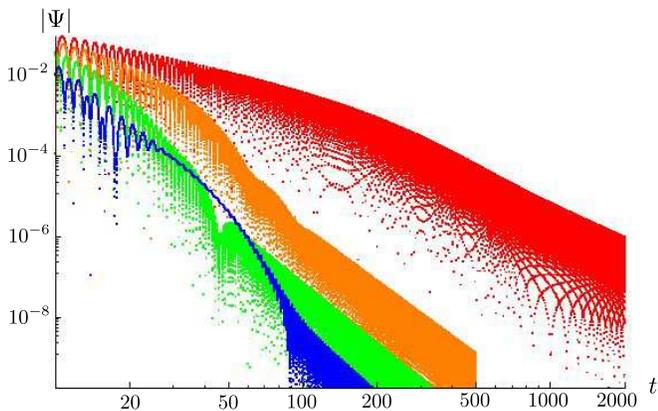}
\caption{Time-domain profiles of black string perturbations for $k=2.5$ $n=2$ (red, top), $n=3$ (orange), $n=4$ (green), $n=5$ (blue, bottom). Late-time decay of perturbations for $n\geq3$ is $\propto t^{-(n+6)/2}$. For lower $n$ the law of decay is different: $\propto t^{-0.93}$ for $n=1$ and $\propto t^{-1.2}$ for $n=2$.}\label{tails}
\end{figure}

An essential advantage of the time domain method in comparison with the Frobenius method is that we do not have any decreasing of the convergence
or loss of accuracy when approaching the point of instability. Therefore our time domain method is more complete than the frequency one, at least  for the dominant mode, which can always be extracted from the time domain picture.
In Fig. 2, one can see examples of time domain profiles for various $n$ and a fixed $k$.
 There one can see that for $n \geq 3$ the intermediate
late time  asymptotic is power law like
$$\Psi \propto t^{-(n+6)/2}, \quad n \geq 3, $$
while for other $n$, the asymptotics are
$$\Psi \propto t^{-0.93}, \quad for \quad  n=1, $$
$$\Psi \propto t^{-1.2} \quad for \quad  n=2.$$
Let us note that this asymptotic apparently should be considered as intermediate. They are expected to go over into other power law ones at very late times, as it takes place for massive fields in general \cite{Simone:1991wn}, \cite{Konoplya:2002wt}, \cite{Ohashi:2004wr}, \cite{Konoplya:2004wg}.

Let us note, that $k$ plays the role of the effective mass. At asymptotically late time we observe power-law damped tails, which have oscillation frequency equal to $k$, resembling asymptotical behavior of massive fields near Schwarzschild black holes. The first overtone's behavior is qualitatively similar to that of the fundamental mode for massive fields of higher-dimensional Schwarzschild black holes \cite{Zhidenko:2006rs}: we can see long-lived oscillations, which can be infinitely long lived modes called the quasi-resonances \cite{Ohashi:2004wr}. The analytical explanation of existence of the quasi-resonances was found in \cite{Konoplya:2004wg}.

\begin{itemize}
\item For $D=5$ ($n=1$), as $k$ grows, the imaginary part of the first overtone quickly decreases and vanishes for some threshold value of $k$, while its real part stays smaller than the threshold value (see Table \ref{n=1.table}). After the threshold value of $k$ is reached, the first overtone ``disappears''.
\item For $D\geq6$ ($n\geq2$), the imaginary part of the first overtone becomes small for large $k$, while the real part asymptotically approaches $k$ (see Fig. \ref{n=2.fig}).
\end{itemize}

 Even though the first overtone of the spherically symmetric black strings behaves similarly to the fundamental mode of massive fields near higher-dimensional Schwarzschild black holes, the other modes have completely different behavior. The fundamental mode of a black string perturbation is purely imaginary. It grows for small values of $k$, leading to instability of the black string. This behavior is common for the unstable modes.

 Indeed, let us multiply the equation (\ref{wavelike}) by the complex conjugated function $\Psi^\star$ and assume that the dependance on time is $\Psi(t,r_\star)=e^{-\imo\omega t}\Psi(r_\star)$. Let us study the integral of the obtained equation
$$I=\intop_{-\infty}^\infty \left(\Psi^\star(r_\star)\frac{d^2\Psi(r_\star)}{dr_\star^2}+\omega^2|\Psi(r_\star)|^2 -V|\Psi(r_\star)|^2\right)dr_\star.$$
 Integration of the first term by parts gives
\begin{eqnarray}\nonumber &&I=
\Psi^\star(r_\star)\frac{d\Psi(r_\star)}{dr_\star}\Biggr|_{-\infty}^\infty + \\\nonumber&&+\intop_{-\infty}^\infty \left(\omega^2|\Psi(r_\star)|^2 -V|\Psi(r_\star)|^2 - \left|\frac{d\Psi(r_\star)}{dr_\star}\right|^2\right)dr_\star = 0.
\end{eqnarray}
Taking into account the boundary conditions (\ref{Frobenius}), we find that the imaginary part of the integral is
\begin{eqnarray}\nonumber Im(I)&=&Re(\sqrt{\omega^2-k^2})|\Psi(\infty)|^2+Re(\omega)|\Psi(-\infty)|^2+
\\\nonumber&&+2Re(\omega)Im(\omega)\intop_{-\infty}^\infty|\Psi(r_\star)|^2dr_\star=0.
\end{eqnarray}
Since the sign of $Re(\sqrt{\omega^2-k^2})$ coincides with the sign of $Re(\omega)$, the non-zero real part of the quasi-normal frequency implies that the imaginary part is negative. Therefore, the unstable modes ($Im(\omega)>0$) must have zero real part.
In other words, \emph{unstable modes cannot be oscillating}.

\begin{figure}
\includegraphics[width=\linewidth]{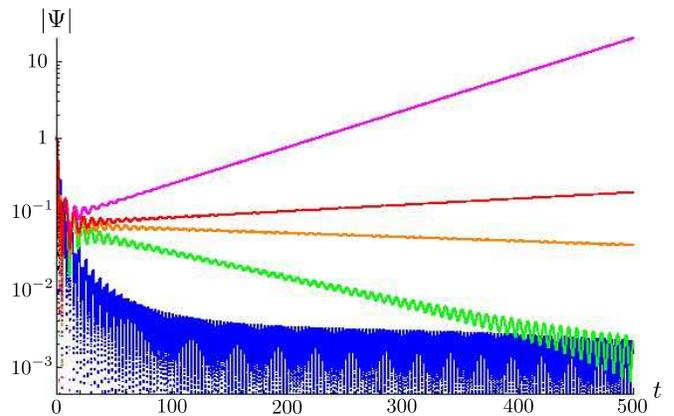}
\caption{Time-domain profiles of black string perturbations for $n=1$ $k=0.84$ (magenta, top), $k=0.87$ (red), $k=0.88$ (orange), $k=0.9$ (green), $k=1.1$ (blue, bottom). We can see two concurrent modes: for large $k$ the oscillating one dominates , near the critical value of $k$ the dominant mode does not oscillate (looks like exponential tail), for unstable values of $k$ the dominant mode grows. The plot is logarithmic, so that straight lines correspond to an exponential decay.}\label{instability}
\end{figure}

\begin{figure*}
\includegraphics[width=.5\linewidth]{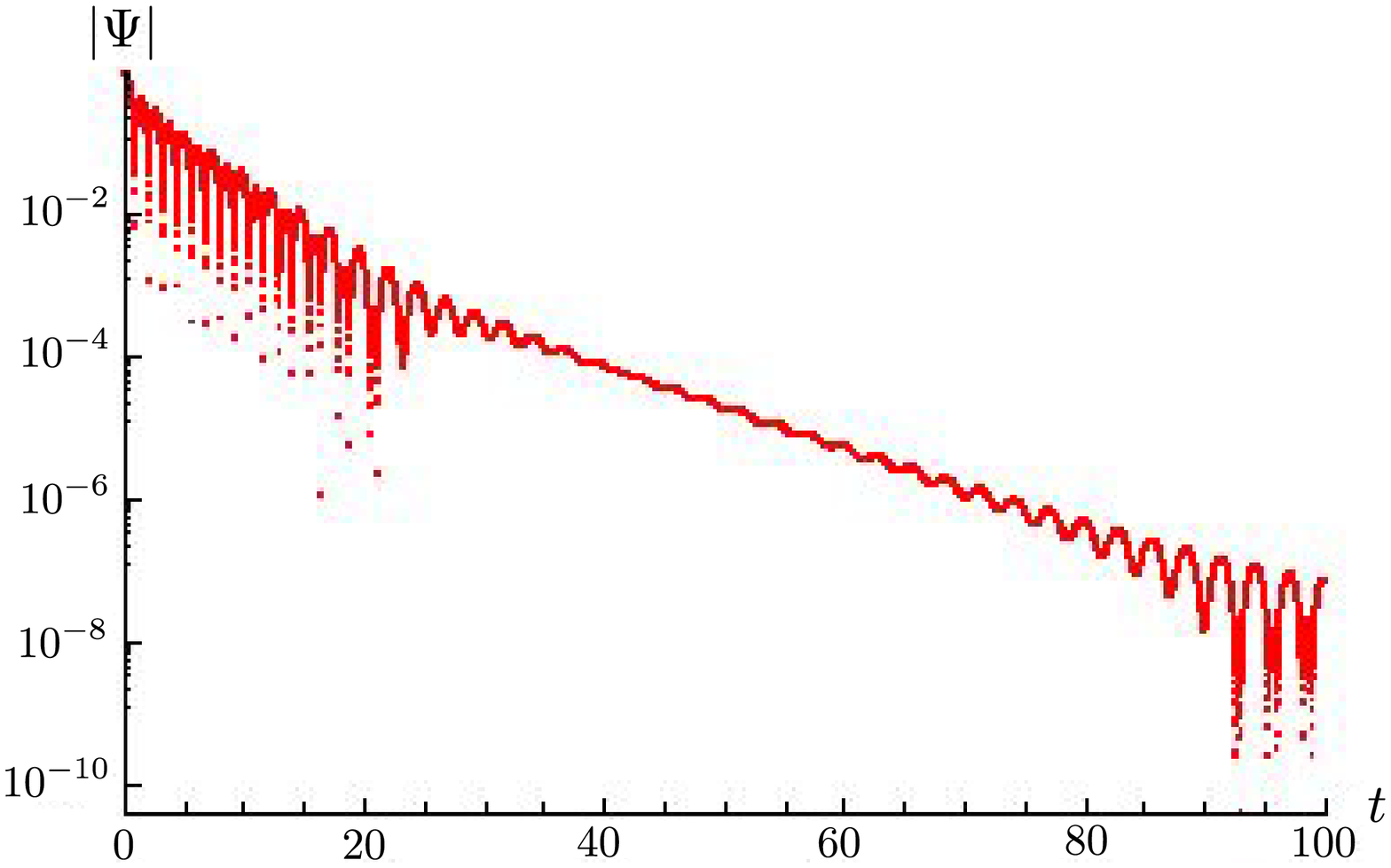}\includegraphics[width=.5\linewidth]{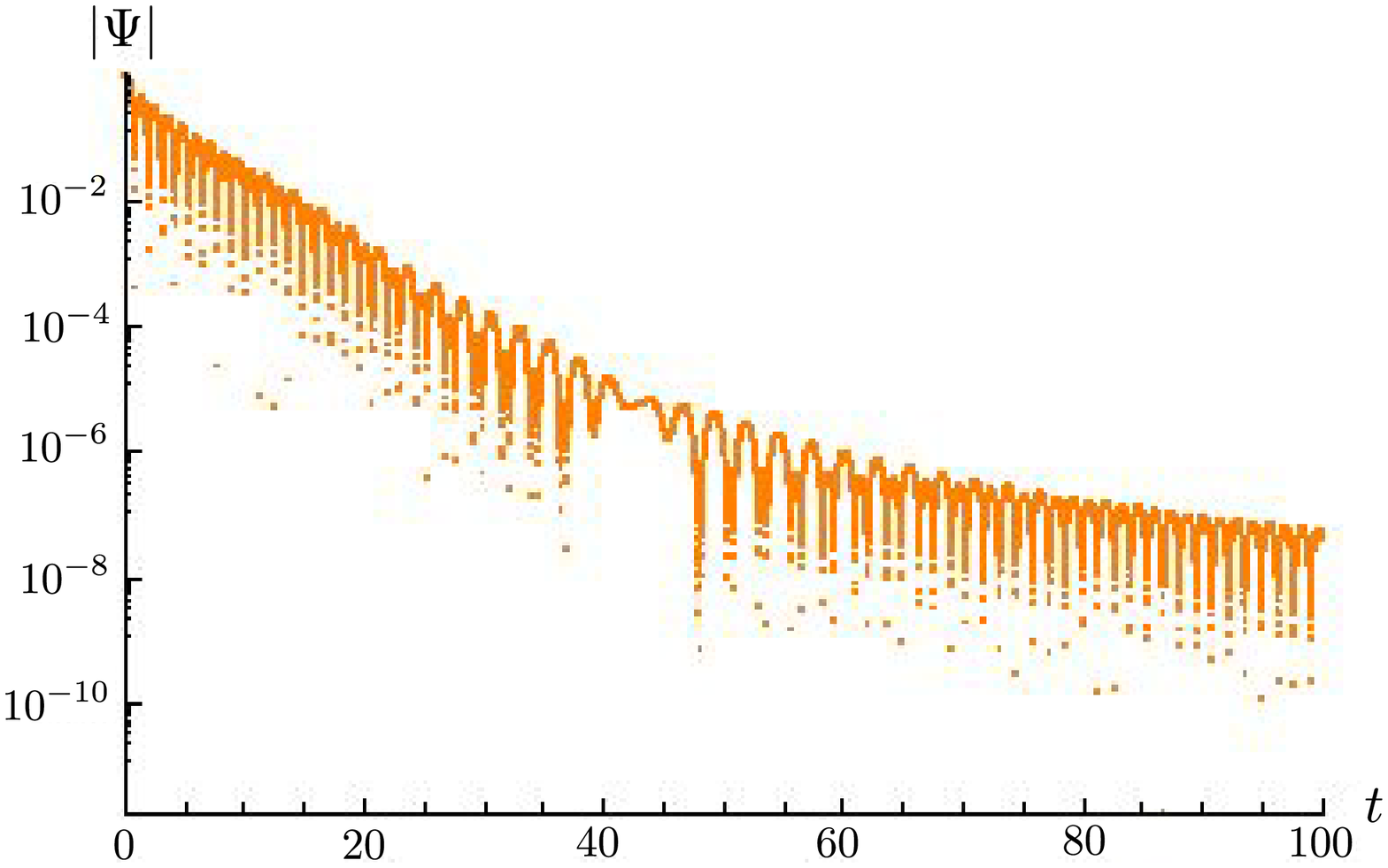}\\\includegraphics[width=.5\linewidth]{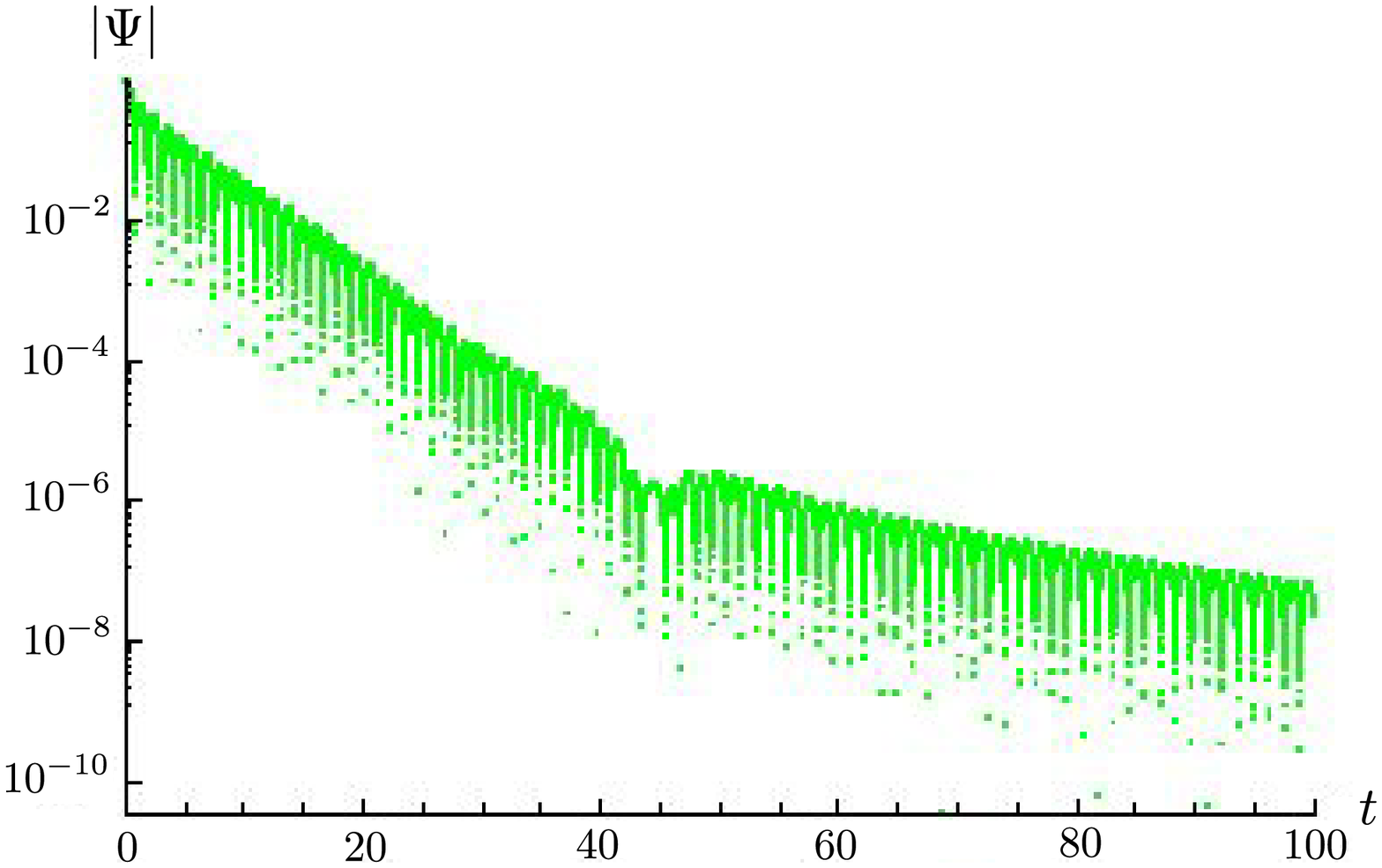}\includegraphics[width=.5\linewidth]{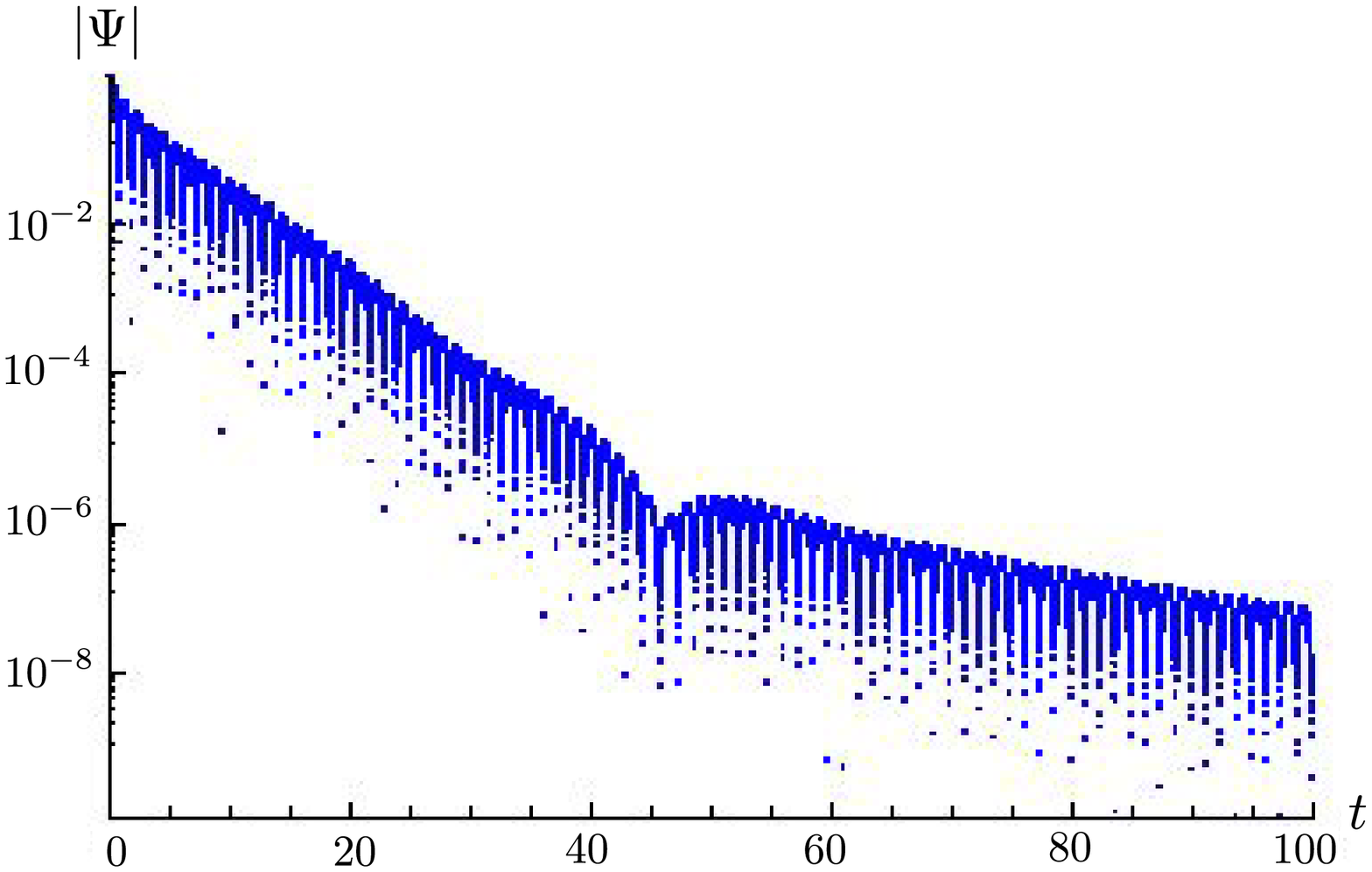}
\caption{Time-domain profiles of black string perturbations for $n=4$ $k=2.2$ (red, left top), $k=2.3$ (orange, right top), $k=2.4$ (green, left bottom), $k=2.5$ (blue, right bottom). Quasi-normal ringing and tails have the same frequency of oscillation which is close to $k$. One can see a period where the oscillation ceases  close  to the critical point.}\label{QNM}
\end{figure*}

 In the stability region, the fundamental mode is also purely imaginary, but damped mode, whose damping rate grows quickly with $k$. Because of its quick growth, this mode cannot be considered as the fundamental one for large $k$. In fact, for larger $k$ (see Tables I-III), the first overtone turns out to be the fundamental mode (the mode with the largest lifetime). The real part of the second overtone decreases, as $k$ grows, and reaches zero for some value of $k$, while the imaginary part remains negative.

Let us now look at the Fig.3. At moderately large values of $k$, sufficiently far from instability, the profile has the same form as that for
massive fields, yet, when approaching the instability point the real oscillation frequency ($Re (\omega)$) and the decay rate
($Im (\omega) < 0$) decrease considerably.
After crossing the instability point we observe, that starting from some tiny values, $Im (\omega) > 0$ are slowly increasing
(while $Re (\omega)$ is still zero for the fundamental mode and tiny oscillations, observed in the time domain, come from the next decayed mode).
Therefore we conclude, that the there is some \emph{static solution} $\omega = 0$ of the wave equation (4), which shows itself \emph{exactly in the threshold point of instability}.
We would say that this picture of instability is natural, if the instability develops on the fundamental mode. However, if instability occurs at higher multipoles
$\ell$, as it takes place for instance in the Gauss-Bonnet theory \cite{ZhidenkoGB}, the picture of instability is quite different: growing modes appear only after rather long period of decaying oscillations (see \cite{ZhidenkoGB}).  Note also, that here we confirmed the threshold values of $k$ found in \cite{Hovdebo:2006jy} with a very good accuracy by the time domain integration (see for instance Fig. 3 for $n=1$). Thus the threshold values are: $k=0.876$ for $n=1$, $k=1.269$ for $n=2$, $k=1.581$ for $n=3$, $k= 1.849$ for $n=4$.

Finally, in Fig. 4, we can see the region of the profiles where the period of the quasinormal ringing goes over into the power-law tail behavior.
Close to the critical point, there exists a period where the oscillation ceases.
This is because the pure damping mode becomes the fundamental mode
near the critical point.

\section{Conclusion}

We have numerically studied the  Gregory-Laflamme instability through quasinormal modes.
 Let us stress three main results obtained here:

1) We have found the quasinormal modes and late-time tails for scalar type of gravitational perturbations of D-dimensional black strings for various $D$, that is for the type of perturbations where the Gregory-Laflamme instability forms in the long wavelength regime.

2) The time domain profiles indicate that the threshold instability value of $k$ corresponds to dominance of some static solution $\omega =0$.

3) Near the instability point (in $k$) the fundamental mode is pure imaginary (non-oscillating), and,  as $k$ is increasing, the lifetime of the second mode is increasing, so that at some moderate $k$ both modes are dominating at the late time of the ringing. At larger $k$, the dominance goes over to the second (oscillating) mode, as to the longer lived one.

Our research could be improved in a number of ways. First of all, one could compute QNMs for higher multipole numbers, starting from the effective potential
derived in \cite{Kudoh:2006bp} and also for other types of gravitational perturbation. Though vector and tensor types of perturbations do not contain instabilities,
such investigation would give us complete data on QNMs and evolution of gravitational perturbations.

The main limitation of our analysis is that
 we cannot say what happen with unstable
black string some time since the moment of initial perturbations: the perturbations will grow and become large, so that the linear approximation
will not be valid anymore. However, it is beyond the scope of this paper.

\begin{acknowledgments}
R. A. K. was supported by the Japan Society for the Promotion of Science
 (JSPS), Japan.
K.M. was supported by JSPS Grant-in-Aid for Scientific Research
No. 19 $\cdot$ 3715.
J.S. was supported by the Japan-U.K. Research Cooperative Program,
  Grant-in-Aid for  Scientific
Research Fund of the Ministry of Education, Science and Culture of Japan
 No.18540262 and No.17340075.
  A. Z. was supported by \emph{Funda\c{c}\~ao de
Amparo \`a Pesquisa do Estado de S\~ao Paulo (FAPESP)}, Brazil.
\end{acknowledgments}

\end{document}